\documentclass[aps,pre,twocolumn,showpacs,floatfix,superscriptaddress,tightenlines,amsmath,amssymb,yhmath]{revtex4-2}
\usepackage{hyperref}
\usepackage[dvipsnames,x11names]{xcolor}
\pagecolor{white}
\usepackage[final]{graphicx}
\usepackage{bm}
\usepackage{float}
\usepackage{cancel}
\usepackage{enumerate, enumitem}
\usepackage[normalem]{ulem}
\usepackage{bibentry}
\usepackage[capitalize]{cleveref}
\crefname{equation}{}{}

\newcommand{\postime}{(\bm{x},t)}

\newcommand{\Tr}{{\rm Tr}}
\newcommand{\Det}{{\rm Det}}
\newcommand{\vel}{\bm{u}}
\newcommand{\pol}{\bm{p}}

\newcommand{\Rey}{{\mathrm{Re}}}

\newcommand{\bdel}{\bm{\nabla}}

\newcommand{\lp}{\lambda_+}
\newcommand{\lm}{\lambda_{-}}

\newcommand{\h}{\bm{h}}

\newcommand{\Sr}{\bm{\Sigma}^{r}}
\newcommand{\Sa}{\bm{\Sigma}^{a}}
\newcommand{\sa}{\sigma_{a}}
\newcommand{\Sij}{\bm{S}}
\newcommand{\Om}{\bm{\Omega}}

\newcommand{\Ep}{\mathcal{E}_{\pol}(q)}

\newcommand{\Eu}{\mathcal{E}_{\vel}(q)}

\begin{document}

\title{Instabilities and turbulence in extensile swimmer suspensions}
\author{Purnima Jain}
\affiliation{Tata Institute of Fundamental Research, Hyderabad 500046, India}
\author{Navdeep Rana}
\affiliation{Max Planck Institute for Dynamics and Self-Organization (MPIDS), D-37077 G\"ottingen, Germany}
\author{Roberto Benzi}
\affiliation{International Research Center for Complexity Sciences, 
Hangzhou International Innovation Institute of Beihang University, Hangzhou, China, 311115}
\affiliation{Department of Physics and INFN, Tor Vergata University of Rome, Via della Ricerca Scientifica 1, 00133 Rome, Italy}
\author{Prasad Perlekar}
\email{perlekar@tifrh.res.in}
\affiliation{Tata Institute of Fundamental Research, Hyderabad 500046, India}

\begin{abstract}
    We study low Reynolds number turbulence in a suspension of polar, extensile, self-propelled inertial swimmers. We review the bend and splay mechanisms that destabilize an ordered flock. The suspension is always unstable to bend perturbations. Using a minimal 1D model, we show that the splay-stable to splay-unstable transition occurs via a supercritical Hopf bifurcation. We perform high-resolution numerical simulations in 2D to study the varieties of turbulence present in this system transitioning from defect turbulence to concentration-wave turbulence depending on a single non-dimensional number, denoting the ratio of the splay-concentration wavespeed to the swimmer motility. 
\end{abstract}

\maketitle

\section{Introduction}
The collective motion of swimmers suspended in a fluid generates complex flows that resemble turbulence at scales much larger than the size of an individual swimmer~\cite{koch2011,marchetti2013,ramaswamy2019,alert2022}. 
Most studies have focused on swimmer suspensions at low Reynolds number ($\Rey \ll 1$) where viscous forces dominate over inertial forces as in microbial suspensions~\cite{simha2002,kruse2004}. In this regime, it is well established that the ordered state of the swimmers is unstable due to the hydrodynamic instability arising from the swimming stresses ($\sa$) exerted by the swimmers on the fluid~\cite{simha2002, ramaswamy2010}. This leads to chaotic flows on a macroscopic scale known as \emph{active turbulence}~\cite{thampi2013, alert2022} that have been observed and investigated in several experimental~\cite{dombrowski2004, wensink2012, dunkel2013, liu2021, peng2021, creppy2015, nishiguchi2015} and numerical studies~\cite{wolgemuth2008,baskaran2009,amin2023, alert2020}. A hallmark feature of these turbulent flows is the presence of topological defects~\cite{marchetti2013,coper2019, thampi2014, Amiri2022, bowick2022}.

In contrast, swimmer suspensions at intermediate Reynolds number ($\Rey \sim \mathcal{O}(1)$), where inertial and viscous forces are comparable, are still less explored~\cite{klotsa2019, chatterjee2021, rana2024, jain2024}. The presence of inertia leads to many interesting and novel behaviors in the collective dynamics of swimmers. In the limit where the swimmer concentration is homogeneous, we have shown that the dimensionless number $R=\rho v_0^2/2 \sa$ (squared ratio of 
the self-advection speed $v_0$ to the invasion speed $\sqrt{\sa/\rho}$ of the instability due to swimming stresses) 
is the relevant control parameter~\cite{chatterjee2021, rana2024}.
For small and intermediate $\Rey$, $\sa$ could be estimated (upto a factor of volume fraction that we take to be of the order of unity~\cite{jain2024}) by viscous stress~\cite{chatterjee2021}, i.e. $\sa \sim \mu v_0/ d$, where $\mu$ is the dynamic viscosity of the fluid, and $d$ is the size of the swimmer. This gives $R \sim \rho v_0 d/\mu = \Rey$ at the scale of a single swimmer.

\begin{figure}[!b]
    \centering
    \includegraphics[width=\linewidth]{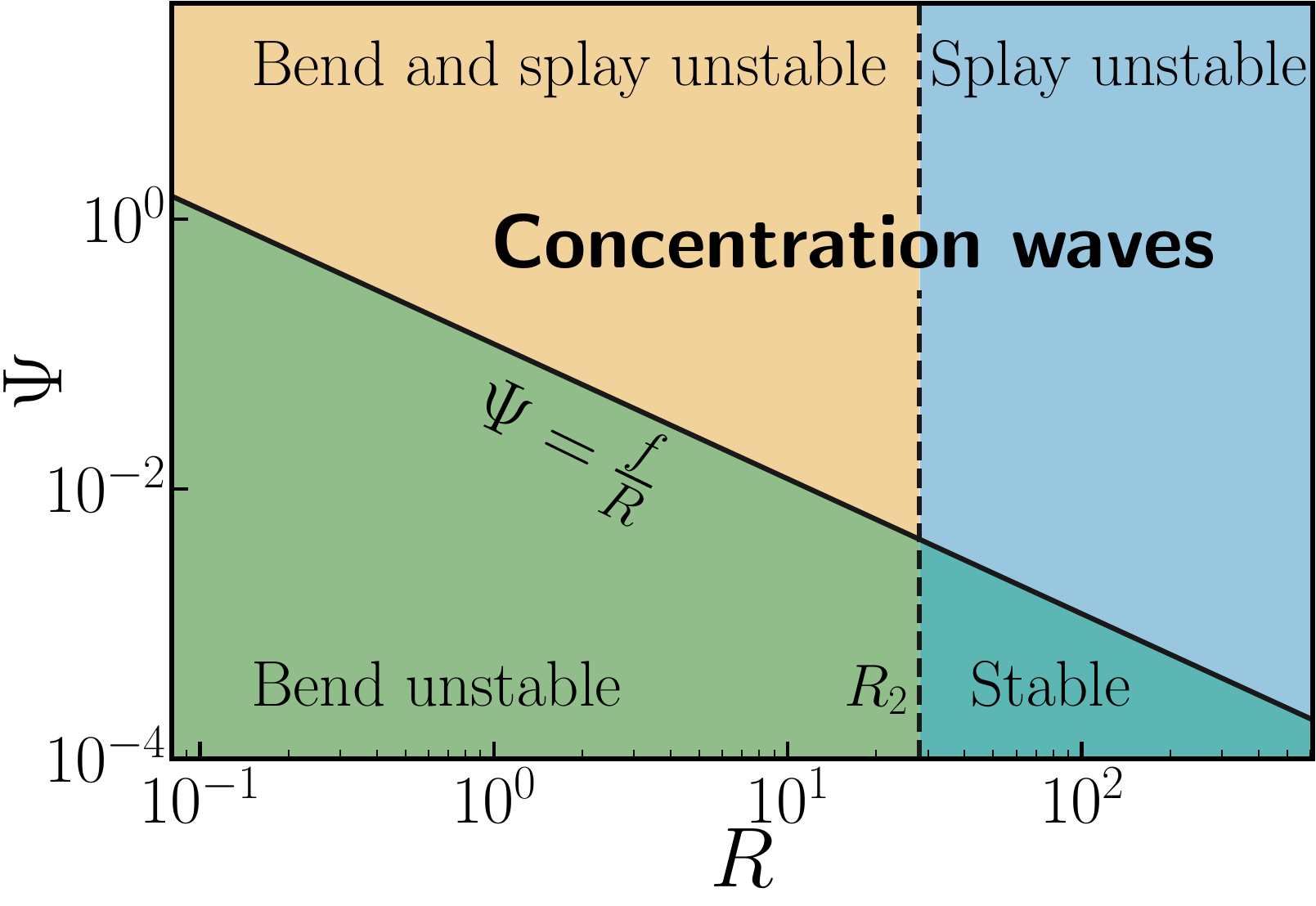}
    \caption{\label{fig:phase-diag} Stability diagram representing the pure bend and pure splay unstable regimes in extensile swimmer suspensions. Figure reproduced from~\cite{jain2024}. The non-dimensional numbers $\Psi \equiv \Gamma E c_0/v_1$ and $R=\rho v_0^2/(2 W c_0)$. }
\end{figure}

At small $R \ll 1$, the polar order parameter exhibits defect turbulence with integer-strength topological defects~\cite{chatterjee2021, rana2024}. Above a threshold $R>R_2$, the ordered state is linearly stable (\cref{fig:phase-diag}). The assumption of homogeneous concentration is justified in the following cases:

\begin{enumerate}
    \item ``Malthusian'' suspensions~\cite{chatterjee2021} where the birth and death processes relax the fluctuations in the concentration of swimmers exponentially fast to the average value so that at the observable time scales, the concentration of swimmers is homogeneous.

    \item Dense suspensions~\cite{rana2024}  where the average value of the concentration of swimmers is so large that fluctuations in the concentration are negligible; therefore the concentration of swimmers is uniform. This renders the polar orientational order parameter divergenceless, $\bdel \cdot \pol=0$.
\end{enumerate}

In general, fluctuations in swimmer concentration cannot be ignored. Therefore, we recently studied the linear stability of an ordered state in this system and discovered a new concentration-wave instability driven by swimmer motility, concentration fluctuations, and inertia~\cite{jain2024}. Interestingly, this instability does not depend on the swimming stresses. Therefore, the consequences of concentration-wave instability on the emerging turbulence were investigated primarily for $R=\infty~ (\sa=0)$ where it was shown that the defect turbulence of the orientational order parameter field at large scales coexists with concentration waves at small scales, which we called concentration-wave turbulence.
We also showed that the concentration-wave turbulence is possible even at small $R$ but for sufficiently large values of 
a dimensionless number 
$\Psi$, 
the squared ratio of the Toner-Tu ``sound'' speed~\cite{toner1998} 
to the self-propulsion speed (\cref{fig:phase-diag}). 
However, the varieties of turbulence accessible for the physically relevant regime of small $R$ but with varying $\Psi$ remains unexplored.

In this work, using high-resolution numerical simulations, we show that for a fixed, small $R=0.1$, as $\Psi$ increases, the steady-state transitions from a defect turbulent state consisting of spiral-asters similar to Malthusian suspensions to an incompressible case with vortical defects, and finally to the novel regime of concentration-wave turbulence. Our study bridges the gap between earlier studies in which limiting cases of homogeneous concentration in polar, active fluids were investigated~\cite{chatterjee2021, rana2024}.

The structure of the paper is as follows. \cref{sec:eqn} provides a brief overview of the hydrodynamic equations in swimmer suspensions, explaining the physical meaning of the relevant terms. In \cref{sec:lsa}, we summarize the results of the linear stability analysis, and in \cref{sec:1dmodel}, using a one-dimensional model that describes the evolution of hydrodynamic fields for pure splay perturbations, we show that the supercritical Hopf bifurcation governs the stability of the system. \cref{sec:simulations} presents the results from direct numerical simulations, where for fixed $R=0.1$, varying $\Psi$ results in transition from defect to concentration-wave turbulence. We characterize this transition using statistical quantities and topological structures in the polar order parameter field. Finally, \cref{sec:conclusions} summarizes and concludes the paper.

\section{Equations \label{sec:eqn}}
The hydrodynamic description of a swimmer suspension based on symmetry and conservation laws yields the following equations~\cite{simha2002, chatterjee2021, giomi2012, rana2024, jain2024},
\begin{align}
    \rho (\partial_t +  \vel \cdot \nabla) \vel&= -\bm{\nabla} \Pi +\mu \nabla^{2}\vel+\bm{\nabla} \cdot \bm{\Sigma}, \label{eq:vel} \\
    \partial_t \pol +(\vel + v_0\pol)\cdot \bm{\nabla} \pol &=  \lambda \bm{S} \cdot \pol + \Om \cdot \pol + \Gamma \bm{h},\mathrm{and} \label{eq:pol} \\
    \partial_t c + \bm{\nabla}\cdot \left[(\vel +v_{1}\pol)c\right] &= D\nabla^2 c \label{eq:conc}.
\end{align}
Equation \cref{eq:vel} describes the conservation of total momentum density $\bm g = \rho \vel$ of swimmers plus fluid, where $\rho$ is the constant suspension mass density and $\vel\postime$ is the suspension velocity. The pressure $\Pi$ imposes the incompressibility constraint $\bdel \cdot \vel=0$. Equation \eqref{eq:pol} describes the dynamics of the polar order parameter $\pol\postime$ which is the local average of the orientation of the swimmers and \eqref{eq:conc} is the continuity equation for the number density of the swimmers $c(\bm{x},t)$.

The scale of the self-propulsion speed $v_1$ is set by the swimming speed of an individual swimmer, and the self-advection speed $v_0$ is the speed at which the polar order parameter advects itself, i.e., transports inhomogeneities in its magnitude and direction~\cite{toner1998, bertin2006, dadhichi2018, dadhichi2020}. These terms are specific to polar active fluids and in general, no symmetry relates $v_0$ and $v_1$~\cite{dadhichi2020}. As in previous works~\cite{giomi2012, alert2022, tjhung2011, cates2018}, to reduce the number of dimensionless parameters we assume $v_0=v_1$ in our numerical investigations, with no discernible impact on their generality. The tensors $\Sij$ and $\Om$ are the symmetric and antisymmetric parts of $\bdel \vel$, and $\lambda$ is the flow alignment parameter~\cite{forster1974, larson1999}. $\Gamma$ is the kinetic coefficient governing the rate of
relaxation of the molecular field $\h$ and is typically expected to be of the order $1/(10\mu)$~\cite{kneppe1983,gennes1993, mazenko1983, jadzyn2001}. The stress $\bm \Sigma \equiv \Sa + \Sr$ consists of the active stress $\Sa =-W c\pol \pol \equiv -\sigma_a\left(c\right) \pol \pol$, arising from the swimming activity, and $\Sr$ is the reversible thermodynamic stress ${\bm \Sigma^r} \equiv \lp \bm{h} \pol + \lm \pol \bm{h}$, with $\lambda_{\pm}\equiv (\lambda\pm 1)/2$~\cite{gennes1993, chandrasekhar1992, chatterjee2021}. The molecular field $\h \equiv -\delta F/ \delta \pol=(a(c)-b|\pol|^2)\pol + K \nabla^2\pol -E \bdel c$ is derived from the free energy functional
\begin{equation}
    F=\int \mathrm{d}{\bm x} \left(-\frac{a(c)}{2}|\pol|^{2}+\frac{b}{4}|\pol|^{4}+\frac{K}{2}|\nabla \pol|^{2}+E \pol \cdot \nabla
    c\right).
\end{equation}
We choose $a(c) = a \tanh(m c)$ with $m>0$, which is consistent with the other forms of $a(c)$ used in literature~\cite{giomi2012, chate2019}. The parameters $a>0$ and $b>0$ ensure the existence of the homogeneous steady state solutions corresponding to the disordered $({\vel}=0,c=c_0,{|\pol|}= 0)$, and an orientationally ordered state $({\vel}=0,c=c_0,{|\pol|}=\sqrt{a(c_0)/b})$. A single Frank's constant $K$ determines the cost of spatial deformation in $\pol$~\cite{frank1958, oseen1933}, and $D$ is the diffusion coefficient of the concentration.

In \eqref{eq:pol}, the term $E \bdel c$ is analogous to the pressure in compressible Navier-Stokes equation~\cite{toner1995, toner1998}. It breaks the $\pol \to -\pol$ symmetry~\cite{kung2006} and is unique to polar fluids. Further, it is present even in equilibrium polar liquid crystals~\cite{kung2006} where it relates to the flexoelectric coupling~\cite{meyer1969}. In the present context of polar active fluids, the coefficient $E$, depending on its sign, promotes alignment or anti-alignment of the orientation of swimmers $\pol$ to the gradients of the concentration $\bdel c$. $E>0$ fosters the movement of swimmers away from high-concentration regions, thus it homogenizes the concentration. $E<0$ promotes the clumping of the swimmers, which could happen for an attractive interaction or through 
Motility-Induced Phase Separation~\cite{cates2015,fily2012athermal,bialke2013,geyer2019}. All the results in the present work pertain to extensile ($W>0$), flow-tumbling ($|\lambda|<1$) swimmers with $E>0$.

\section{Linear stability analysis\label{sec:lsa}}

In \cref{fig:phase-diag}, we reproduce the stability diagram from~\citet{jain2024} that shows different unstable regimes in extensile swimmer suspensions. We summarize here the main results for the convenience of the reader; details are available in the supplemental material of \cite{jain2024}.

The two primary deformations of the polar order parameter are the pure bend and the pure splay modes. For pure bend modes, the perturbations vary along the ordering direction ($\hat{\bm x})$, therefore, the perturbation wavemodes are $\bm q = q \hat{\bm x}$.
In this case, concentration fluctuations decouple from the polar order parameter and the hydrodynamic velocity fields in the linearized equations, resulting in a bend instability of the ordered state for $R<R_2$. Here, $R_2 = \lp(1+\beta)^2/2\beta$, with $\beta=\Gamma K\rho/\mu$, and $R=\rho v_0^2/(2 W c_0)=\rho v_0^2/2\sa(c_0)$. The ordered state for bend perturbations is stable for $R>R_2$ \cite{chatterjee2021, rana2024, jain2024}.

For pure splay modes, the perturbations vary transversally to the ordering direction ($\hat{\bm x})$, therefore, the perturbation wavemodes are $\bm q = q \hat{\bm y}$. This yields splay instability of the ordered state if
\begin{equation}\label{eq:threshold_E}
    \Psi \equiv \frac{\Gamma E c_0}{v_1} > \frac{f}{R},
\end{equation}
with
\begin{equation}
    f=\left(\frac{v_0^2}{2 v_1^2}\right)\left(\frac{|\lm| p_0^2\Gamma \mu(1 + \beta)}{\lm^2 p_0^2-\Gamma \rho (D+\Gamma K)}\right).
\end{equation}
It is evident from \eqref{eq:threshold_E} that the splay instability or concentration-wave instability exists even when active stress is zero ($R=\infty$). This case was explored in detail in~\cite{jain2024}. In the next section, we discuss this instability for fixed $R=0.1$ using a minimal 1D model.

\section{Splay stable-unstable transition in a minimal 1-D model\label{sec:1dmodel}}

We now construct a minimal 1D model, where the relevant terms in the linear stability analysis of splay perturbations, and the cubic damping in \eqref{eq:pol} are retained. The equations are,
\begin{align} \label{eq:1dmodel}
    \begin{split}
        \rho \partial_t u &= \mu \partial_y^2 u + \lambda_- p_0 E \partial_y^2 c - W p_0 \partial_y (pc),\\
        \partial_t p &= \lambda_- p_0 \partial_y u -\Gamma b p^3 + \Gamma K \partial_y^2 p - \Gamma E \partial_y c,~\text{and} \\
        \partial_t c &= -v_1 \partial_y(pc) + D \partial_y^2 c.
    \end{split}
\end{align}
As expected, the above system correctly captures the splay stable-unstable transition as shown in \cref{fig:phase-diag}. Consider the following ansatz for \eqref{eq:1dmodel},
\begin{align}\label{eq:ansatz}
    \begin{split}
        u(y,t) &= U(t) \sin(qy),\\
        p(y,t) &= P(t) \cos(qy),~\text{and}\\
        c(y,t) &= c_0 + C(t) \sin(qy).
    \end{split}
\end{align}

Substituting \cref{eq:ansatz} in \cref{eq:1dmodel} and using the method of harmonic balance~\cite{deuflhard2011, mickens1984}, we obtain the amplitude equations for a single wave number $q$ within the minimal 1D model, which are,
\begin{align} \label{eq:amplitude}
    \begin{split}
        \rho \dot{U} &= -\mu q^2 U - \lambda_- p_0 E q^2 C + W p_0 c_0 q P,\\
        \dot{P} &= \lambda_- p_0 q U -\frac{3 \Gamma b}{4} P^3 - \Gamma K q^2 P - \Gamma E q C,~\text{and} \\
        \dot{C} &= v_1 c_0 q P - D q^2 C.
    \end{split}
\end{align}

The linear stability criterion for \cref{eq:amplitude} about the fixed point ($U=P=C=0$) is identical to \cref{eq:threshold_E}. The region corresponding to $\Psi>f/R$ is linearly unstable, and $\Psi<f/R$ is linearly stable.
\begin{figure}[h!]
    \centering
    \includegraphics[width=\linewidth]{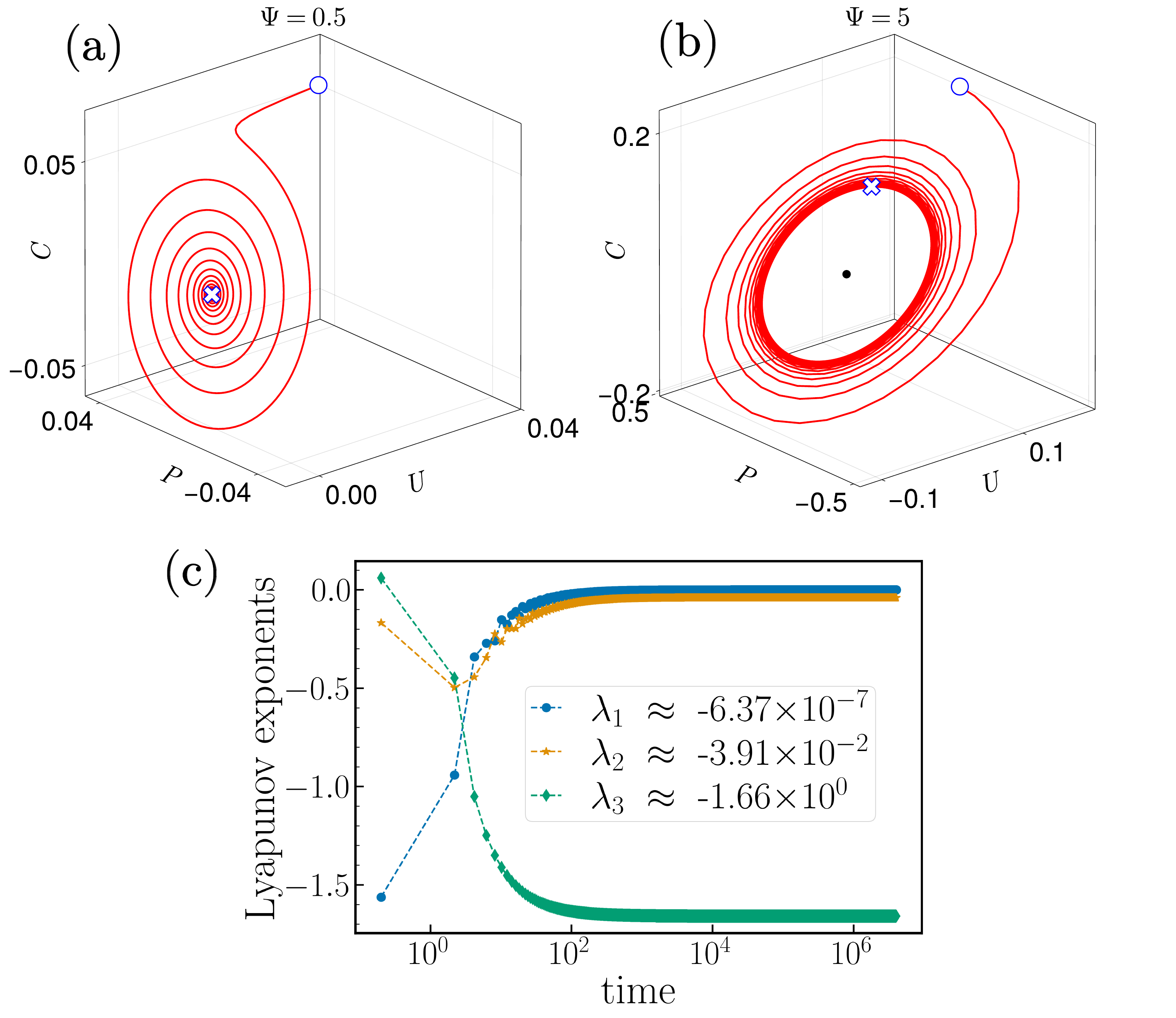}
    \caption{\label{fig:1d_stability} Representative trajectories in the ($U,P,C$) space for (a) $\Psi=0.5$ and (b) $\Psi=5$. The initial condition is marked with an open circle, and the final state is marked with a cross. 
    The system undergoes a supercritical Hopf bifurcation at $\Psi=f/R$. (a) For $\Psi <f/R$, the system is linearly stable and the initial perturbation spirals inward to the fixed point at the origin. (b) For $\Psi >f/R$, the system is linearly unstable and the initial perturbation settles down in a limit cycle around the fixed point. (c) shows the time evolution of the Lyapunov exponents for $\Psi=5$. As expected, for a limit cycle, the exponents converge to a near-zero ($\lambda_1 \approx -6.37 \times 10^{-7}$), and two negative values ($\lambda_2 \approx -3.91 \times 10^{-2}$, $\lambda_3 \approx -1.66$). 
    }
\end{figure}

Numerical integration of the amplitude equations \cref{eq:amplitude} for the parameters $\rho=1$, $\mu=0.1$, $W=0.05$, $\Gamma = 1$, $D=10^{-4}$, $K=10^{-3}$, $\lambda=0.1$, $v_1=v_0=0.1$, $b=1$, $c_0=1$, $p_0=\sqrt{0.1}$, and $q=4$ reveals the following picture. For $\Psi<f/R$, the solution trajectories spiral inward in a plane towards the stable fixed point. In contrast, for $\Psi> f/R$, we find that the long-time dynamics is still confined in a plane, however, the trajectories now make a closed orbit, indicating the birth of a limit cycle. This suggests that the stable-unstable transition in \cref{eq:amplitude} occurs via a supercritical Hopf bifurcation at the threshold $\Psi=f/R$. The phase portrait of the corresponding trajectories is shown in \cref{fig:1d_stability}. To further verify that the orbit corresponds to a limit cycle, we evaluate the Lyapunov exponents of \eqref{eq:amplitude} with $\Psi=5$ following the procedure described in~\cite{wolf1985}. For a dissipative system, unless the dynamics ends at a fixed point, one of the Lyapunov exponents vanishes~\cite{haken1983}. Consistent with the presence of a limit cycle, we find two negative and one near-zero ($\sim 10^{-7}$) Lyapunov exponent (see \cref{fig:1d_stability}(c)).

\section{Statistically steady states\label{sec:simulations}}
We now systematically investigate the statistically steady states obtained for different values of $\Psi$ for a fixed $R=0.1$. In particular, we monitor the transition from defect turbulence to concentration-wave turbulence with changing $\Psi$.

\subsection{Direct Numerical Simulations}

\begin{figure*}[b]
    \centering \includegraphics[width=\linewidth]{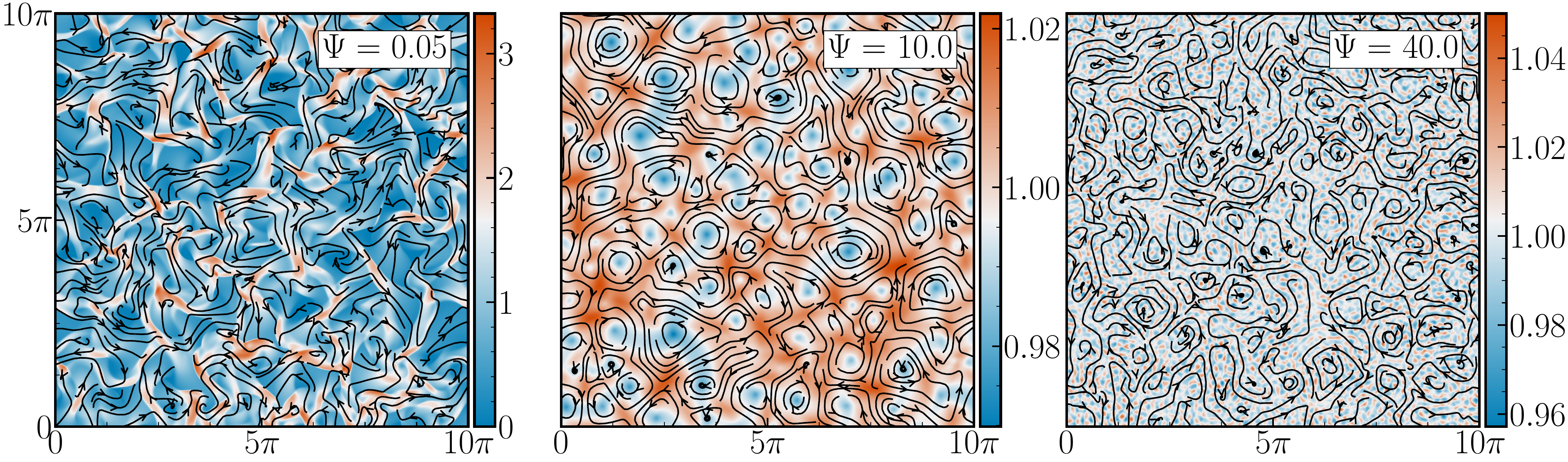}
    \caption{\label{fig:snaps} Pseudocolor plot of the concentration field with streamlines of the polar order parameter for $R=0.1$. As $\Psi$ increases, the fluctuations in the concentration reduce and the topological structures change from spiral-asters defects at $\Psi=0.05$ to vortices at $\Psi=10$. For $\Psi=40$, the concentration-wave instability becomes dominant, giving rise to concentration-wave turbulence.}
\end{figure*}

We perform simulations in a square domain of length $L=10\pi$ discretized with $N^2=1024^2$ collocation points. We use a hybrid numerical integration scheme that uses pseudo-spectral method to integrate \cref{eq:vel} and fourth-order centered finite difference to calculate spatial derivatives in \cref{eq:pol} and \cref{eq:conc}. For time integration, we use a second-order Adams-Bashforth scheme.
The parameters $\rho=1$, $\mu=0.1$, $W=0.05$, $\Gamma = 1$, $D=10^{-4}$, $K=10^{-3}$, $\lambda=0.1$, $v_1=v_0=0.1$, $a=0.1$, $b=1$, $m=10$, and $c_0=1$ are kept fixed. This fixes $R=0.1$, $\beta=10^{-4}$ and we vary $\Psi$ by changing $E$. 

In \cref{fig:snaps}, we show steady-state snapshots of the concentration field overlaid with the streamlines of the polar order parameter. For $\Psi=0.5$, the system is bend unstable and we observe defect turbulence with spiral-aster-like defects of the polar order parameter. This resembles defect turbulence in Malthusian suspensions, where the defects are asters~\cite{chatterjee2021}. As $\Psi$ increases, around $\Psi=10$, the aster-like defects transition to vortices. These structures are similar to those observed in dense suspensions where the polar order parameter is incompressible~\cite{rana2024, rana2020}. Finally, for large values of $\Psi$, we observe turbulence triggered by the concentration-wave instability~\cite{jain2024}.

\subsection{Compressibility via divergence of the polar order parameter}
\begin{figure*}[ht!]
    \centering
\includegraphics[width=0.79\linewidth]{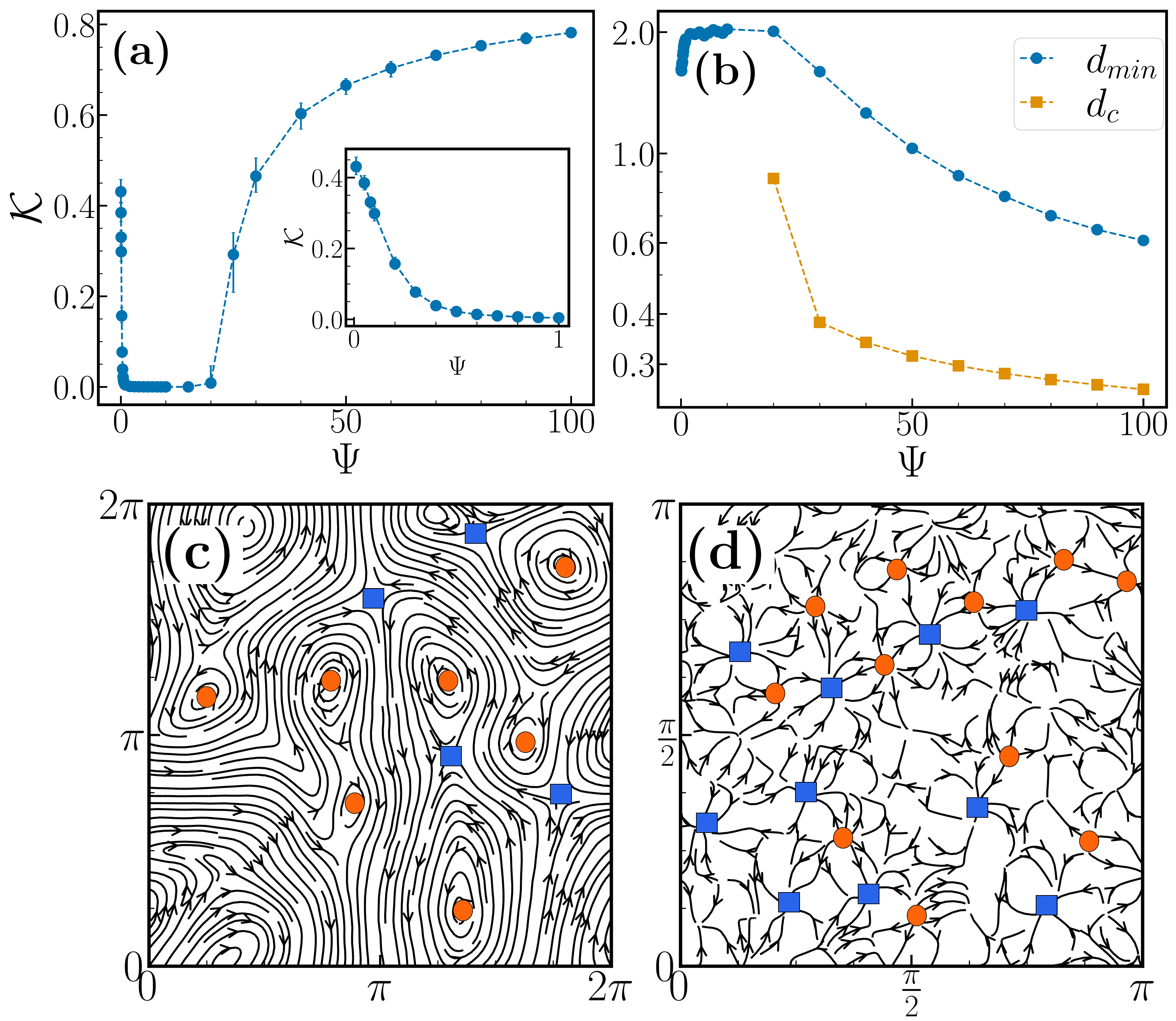}
    \caption{\label{fig:comprs} (a) Variation of the compressibility ($\mathcal{K}$) of the polar order parameter with $\Psi$. The system transitions from compressible to incompressible, and again to a compressible regime when the concentration-wave instability dominates over the bend instability. Inset: Zoomed in plot showing decrease of $\mathcal{K}$ at small $\Psi$. (b) Plot of average distance between nearest neighbor defects for full $\pol$ ($d_{min}$), and the compressible part $\pol_c$ ($d_c$) in the concentration-wave turbulent regime. $d_{min}$ shows an opposite trend to $\mathcal{K}$: It first increases in the region where the system is compressible, then becomes nearly constant in the incompressible region, and finally, it again decreases where $\mathcal{K}$ increases. The inter-aster separation $d_c$ decreases with $\Psi$. (c) and (d) show the streamlines of the solenoidal ${\bm p}^{i}$ and the potential ${\bm p}^c$ components of ${\bm p}$ for $\psi=40$, respectively. Some of the defects are shown in (c), consisting of vortices (orange circles) and saddles (blue squares), and in (d), consisting of outward asters (orange circles) and inward asters (blue squares). Subdomains of $L=2\pi$ and $L=\pi$ are shown respectively for clarity.}
\end{figure*}

To characterize the different types of structures (asters, spirals, vortices, and waves) observed, we compute the scaled variance of $\nabla \cdot \pol$ as a measure of compressibility, 
\begin{align}
    \mathcal{K} = \langle (\bm{\nabla \cdot \pol})^2\rangle/\langle|\bm{\nabla} \pol|^2\rangle,
\end{align}
where $\langle \cdot \rangle$ denotes the spatio-temporal average in the steady state. For an isolated aster $\mathcal{K}=1$, while for an isolated vortex $\mathcal{K}=0$. The plot of $\mathcal{K}$ versus $\Psi$ is shown in \cref{fig:comprs}(a). With increasing $\Psi$, the structures in the polar order parameter transition from spiral-asters to vortices (see \cref{fig:snaps}). This explains the initial decrease in $\mathcal{K}$ with increasing $\Psi$. However, $\mathcal{K}$ increases beyond $\Psi=1$ which corresponds to the concentration-wave turbulence regime.

To understand the increase in compressibility, we consider the field ${\bm p}$ for $\Psi=40$ (see \cref{fig:snaps}). Using Helhmoltz decomposition, we resolve ${\bm p}= {\bm p}^{c} + {\bm p}^{i}$ into a potential ${\bm p}^c=-\nabla \phi$ and a solenoidal field ${\bm p}^i$. A snapshot of ${\bm p}^i$ and ${\bm p}^c$ is shown in \cref{fig:comprs}(c and d).  
It is apparent from \cref{fig:snaps} ($\Psi=40$) that ${\bm p}$ consists of large-scale vortical structures and small-scale asters. We have verified that the inter-aster separation $d_c$ decreases with increasing $\Psi$. These asters cause an increase in compressibility, which increases as inter-aster separation decreases (see \cref{fig:comprs}(b)). 

\subsection{Local structures of the polar order parameter}
\begin{figure}[h!]
    \centering
    \includegraphics[width=0.9\linewidth]{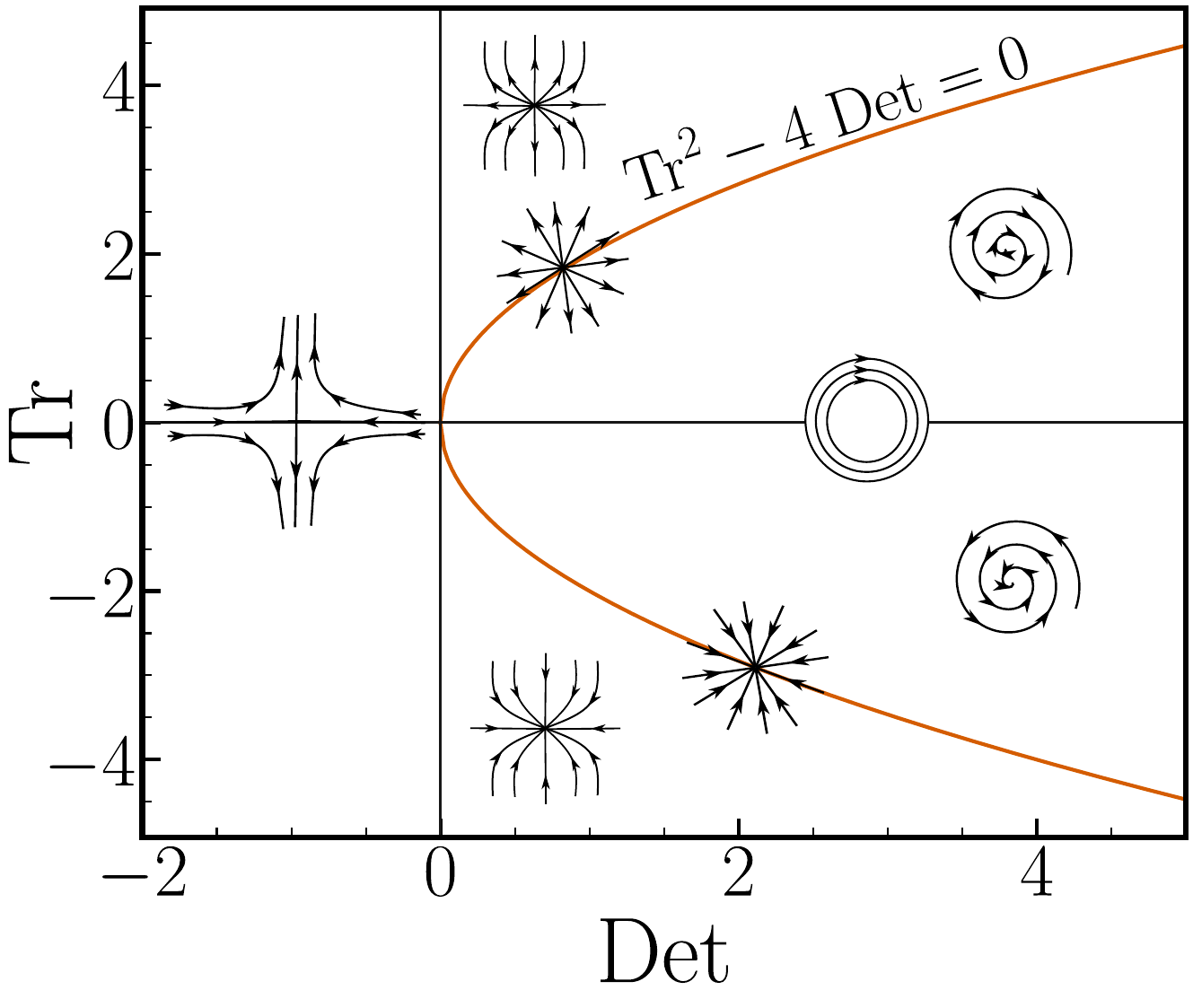}
    \caption{\label{fig:schematic} Schematic showing possible local structures in polar order parameter based on the eigenvalues of the $\nabla {\bm p}$ tensor in the Det-Tr plane. The parabola $\rm P\equiv \Tr^2-4~\Det=0$ separates regions with real and complex eigenvalues. The different structures are (a) $\rm P>0$: Saddles, (b) $\rm P<0$: Spirals and centers ($\Tr=0$), and (c)  $\rm P=0$: Star nodes or asters.}
\end{figure}

As discussed above, the streamlines of the field ${\bm p}$ show a variety of structures. The trace (\Tr) and the determinant (\Det) invariants of the $\nabla \pol$ tensor provide a natural way to characterize the topology of the structures~\cite{strogatz2015}. A schematic showing possible structures in the $\Tr$-$\Det$ plane is shown in \cref{fig:schematic}. The parabola ${\rm P}\equiv {\Tr}^2 - 4~{\Det}$, delimitates different configurations: (a) Outward-pointing star node (${\rm P}=0$,  $\Tr>0$), (b) Outward-pointing spiral (${\rm P}<0$, $\Tr>0$), (c) a vortex (${\rm P}<0$, $\Tr=0$), and (d) a saddle (${\rm P}>0$). Identical but inward-pointing configurations to (a) and (b) occur for ${\rm P} \leq 0$, $\Tr<0$.

\begin{figure}[h!]
    \centering
    \includegraphics[width=\linewidth]{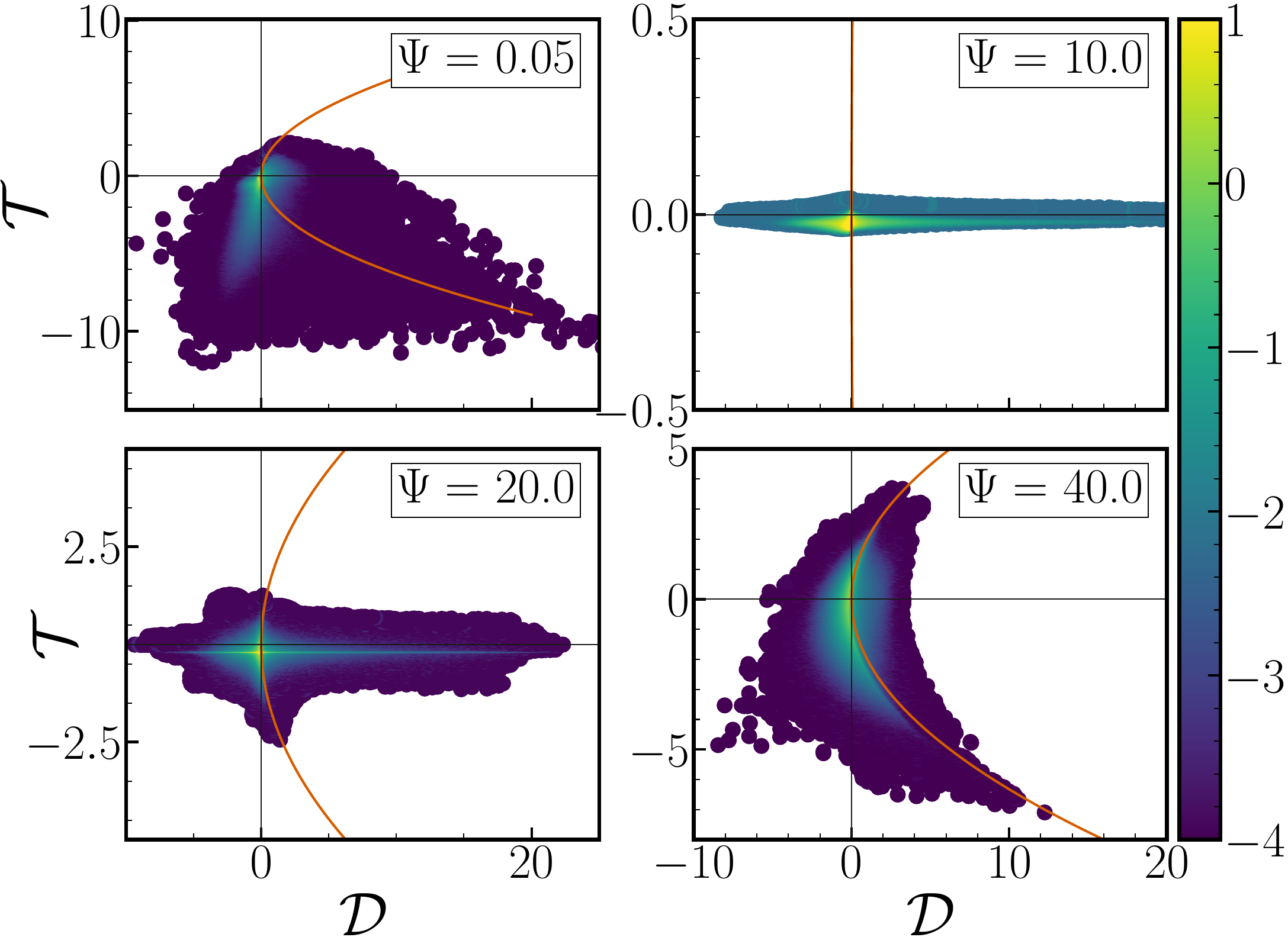}
    \caption{\label{fig:R0.1_jointpdf} Joint PDF $Q({\mathcal T},{\mathcal D})$ for $R=0.1$ and varying $\Psi$. Here, $\mathcal{T}=\Tr/\sqrt{\langle P_{ij}P_{ij}} \rangle$, $\mathcal{D}=\Det/\langle P_{ij}P_{ij} \rangle$, with $P_{ij}= \partial_i p_j + \partial_j p_i$. For small $\Psi=0.05$, the structures consist mainly of inward-pointing asters and saddles, similar to Malthusian suspensions \cite{chatterjee2021}. For intermediate $\Psi=10$, $Q$ lies along $\mathcal{T}=0$, indicating vortices and saddles, similar to the incompressible dense suspensions \cite{rana2024}. For $\Psi=20$, transition to concentration-wave turbulence regime \cite{jain2024} begins where $Q$ starts to broaden along the parabola, indicating the appearance of asters. Finally, for large $\Psi=40$, we observe that $Q$ mainly spreads along the parabola, indicating the presence of a large number of asters. 
}
\end{figure}

In \cref{fig:R0.1_jointpdf}, we plot the joint PDF $Q(\mathcal{T},{\mathcal{D}})$ for different values of $\Psi$, where $\mathcal{T}=\Tr/\sqrt{\langle P_{ij}P_{ij}} \rangle$, $\mathcal{D}=\Det/\langle P_{ij}P_{ij} \rangle$, $P_{ij}= \partial_i p_j + \partial_j p_i$, and repeated indices are summed over. The shape of the PDFs indicates changes in the structures in the polar order parameter with increasing $\Psi$. For small $\Psi=0.05$, the structures consist mainly of inward-pointing spirals and saddles. At intermediate $\Psi=10,~20$, we observe that the PDFs are mostly concentrated along the $\mathcal{T}=0$, indicating that the structures consists primarily of vortices and saddles. This is consistent with our observation in~\cref{fig:comprs} that $\pol$ is nearly incompressible ($\mathcal{T} \ll 1$) for these values of $\Psi$. Finally, for large values of $\Psi=40$, the local structures are dominated by asters. This is also consistent with our observation that the large compressibility ${\mathcal{K}}$ at $\Psi=40$, is due to the presence of asters in the compressible part of the polar order parameter. 

\clearpage

\subsection{Energy spectrum}
In this section, we discuss the shell-averaged energy spectrum for polar order parameter and velocity fields to characterize different turbulent regimes depending on the values of $\Psi$. The energy spectrum is defined as,
\begin{align}
    \Ep \equiv \sum'_{\bm{k}} |\hat{\bm \pol}_{\bm k}|^2,~\mathrm{and},~\Eu \equiv \sum'_{\bm{k}} |\hat{\vel}_{\bm k}|^2,
\end{align}
where $\hat{()}_{\bm k}$ denotes the Fourier coefficient of the wavenumber $\bm k$ and the primed sum is defined as $\sum'_{\bm{k}} = \sum_{q - \pi/L \leq |\bm k| < q + \pi/L}$. 

\subsubsection{Polar order parameter}

\begin{figure}
    \centering
    \includegraphics[width=\linewidth]{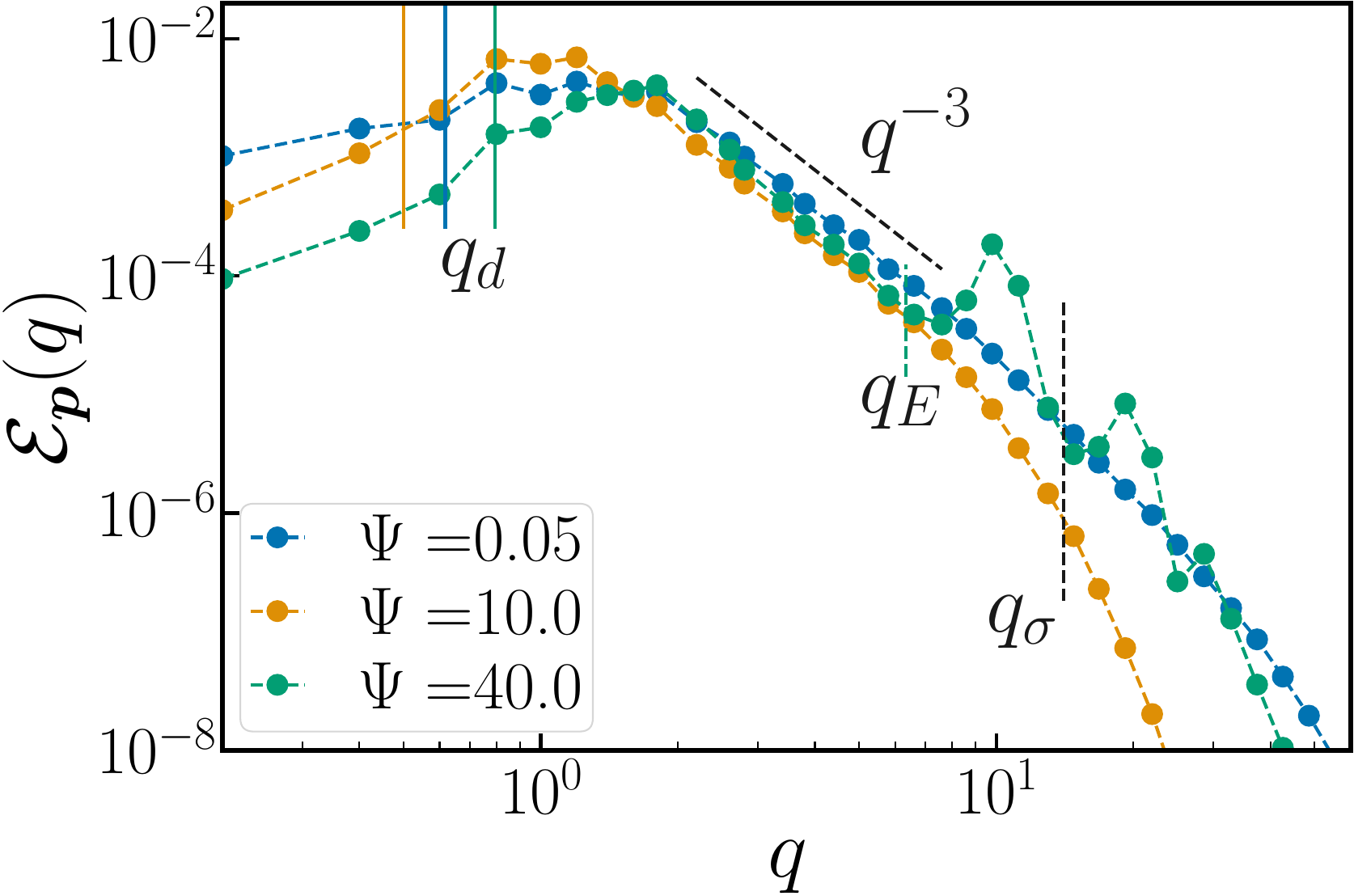}
    \caption{ \label{fig:Ep} Plot of the polar order parameter spectrum for different values of $\Psi$. The spectrum for all the values peaks at around $q \sim q_d$ where $q_d = 1/d_{min}$, shown with vertical lines of the same color.
    The spectrum shows Porod's scaling ($\sim q^{-3}$) up to $q \sim q_\sigma$ in the defect turbulent regime ($\Psi=0.05,~10$) and $q \sim q_E$ in the concentration-wave turbulent regime ($\Psi=40$).}
\end{figure}

In \cref{fig:Ep}, the polar order parameter spectrum $\Ep$ is shown for different values of $\Psi$. The spectrum peaks around $q \sim q_d$, where $q_d = 1/d_{min}$, the inverse of the average distance between the nearest-neighbor defects. The spectrum shows Porod's scaling $\Ep \sim q^{-3}$ for all $\Psi$. In the defect turbulent regime ($\Psi=0.05,~10$), the scaling follows up to $q \sim q_\sigma$ where $q_\sigma = 2\pi/\ell_\sigma = \mu/\sqrt{\rho W c_0}$. This is consistent with the previous results for dense suspensions~\cite{rana2024}. In the concentration-wave turbulent regime ($\Psi=40$), Porod scaling follows up to $q<q_E$, where $q_E = 1/\ell_E = 1/\sqrt{(\Gamma E v_1-W \lm p_0^2/\rho)c_0}$, which is the dominant wavenumber of the concentration waves~\cite{jain2024}.

\subsubsection{Hydrodynamic velocity}

\begin{figure*}
    \centering
    \includegraphics[width=\linewidth]{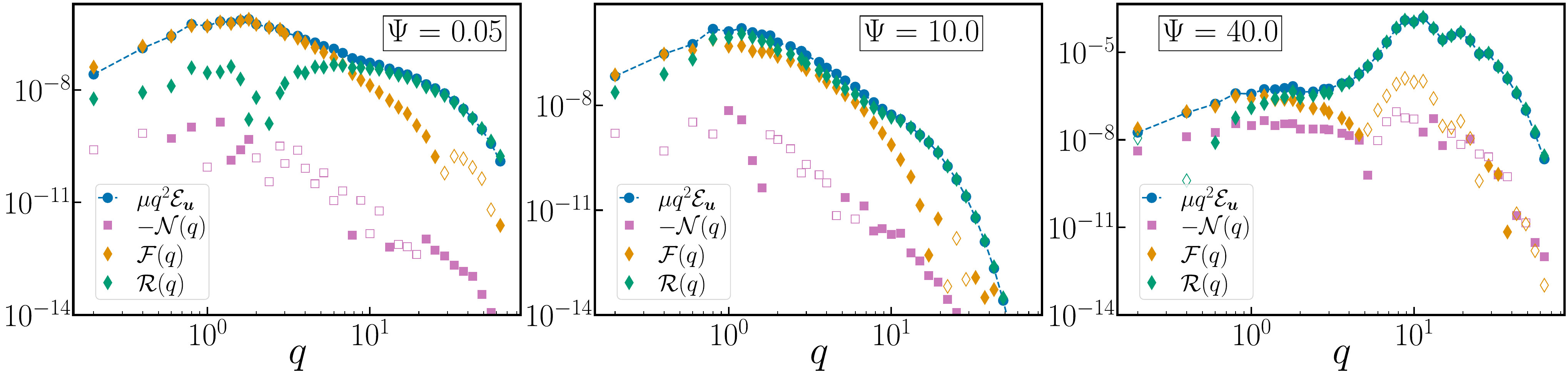}
    \caption{\label{fig:balance} Kinetic energy spectrum for varying $\Psi$. For $\Psi=0.05$, the active stress balances the viscous dissipation at large scales. For $\Psi=10$, where the system is nearly incompressible, both the active and reversible stress are important. Finally, for $\Psi=40$ corresponding to the concentration-wave turbulence regime, reversible stresses balance viscous dissipation consistent with \cite{jain2024}. The open markers denote the negative value of the corresponding quantity. }
\end{figure*}

We now characterize the kinetic energy spectrum $\Eu$ in different turbulent regimes by studying the energy balance that reads,
\begin{align}\label{eq:balance}
    {\mu q^2 \Eu \approx - \mathcal{N}}(q) +  {\mathcal{F}}(q) + \mathcal{R}(q),
\end{align}
where the nonlinear transfer ${\mathcal{N}}(q) = \rho \sum_{\bm k}^\prime \mathrm{Re}[\hat{\vel}(-{\bm k}) \cdot (\mathcal{P}\cdot\widehat{[\vel \cdot \nabla \vel]}({\bm k})]$, the contribution due to the active stress ${\mathcal{F}}(q) = \sum_{\bm k}^\prime \mathrm{Re}[\hat{\vel}(-\bm k) \cdot (\mathcal P \cdot (i \bm{k}\cdot\hat{\bm \Sigma}^a({\bm k}))]$, the contribution due to the restoring stress ${\mathcal{R}}(q) = \sum_{\bm k}^\prime \mathrm{Re}[\hat{\vel}(-\bm k) \cdot (\mathcal P \cdot (i \bm{k}\cdot\hat{\bm \Sigma}^r({\bm k}))]$, and $\mathcal{P}={\mathcal I}- {\bm q}{\bm q}/q^2$ is the projection operator.

In \cref{fig:balance}, we show the energy balance for different values of $\Psi$. For small $\Psi=0.05$, characteristic of the Malthusian regime, viscous stress is mainly balanced by active stresses in the low-wavenumber range ($q<q_\sigma$), where $\Ep$ exhibits Porod scaling (\cref{fig:Ep}). At higher wavenumbers ($q>q_\sigma$), viscous dissipation is balanced by the restoring stress. In contrast, for large $\Psi=40$, corresponding to the concentration-wave turbulence regime, the restoring stress predominantly balances viscous dissipation at $q>1$, consistent with the findings of \cite{jain2024}. At low wavenumbers ($q<1$), active stresses take over the balance. Interestingly, in the intermediate regime ($\Psi=10$), although the defect structures are similar to those of incompressible dense suspensions \cite{rana2024}, the energy balance differs. In our case, both active and restoring stresses contribute to the balance of viscous dissipation, in contrast to dense suspensions \cite{rana2024}, where only active stresses play this role.

\section{Conclusions\label{sec:conclusions}}
In this paper, we analyze in detail the regime $R=0.1$ corresponding to polar, extensile swimmer suspensions where two instabilities, namely bend and concentration-wave, arise (\cref{fig:phase-diag}) depending on a dimensionless number $\Psi$. Using a minimal 1D model, we showed that the latter instability is governed by a supercritical Hopf bifurcation.

We further study the nonequilibrium steady states in two dimensions that arise as the system transitions from a defect turbulent regime at small $\Psi$ to a concentration-wave turbulent regime at large $\Psi$. This transition gives rise to an increase in the compressibility of the concentration, which we have defined in terms of the polar order parameter due to the emergence of aster defects in the concentration-wave turbulent regime. We also study the topological structures and energy spectrum in the polar order parameter field, along with the energy balance of hydrodynamic velocity in different regimes of turbulence. 

Our study provides a comprehensive understanding of the instabilities and turbulence in inertial active suspensions of extensile swimmers unifying the earlier limiting cases with homogeneous concentration~\cite{chatterjee2021, rana2024}.

\section*{Acknowledgements}
We thank Sriram Ramaswamy for the critical reading of the paper and useful discussions. PJ and PP acknowledge support from the Department of Atomic Energy (DAE), India under Project Identification No. RTI 4007, and DST (India) Project Nos. MTR/2022/000867. All the simulations are performed using the HPC facility at TIFR Hyderabad.

%

\end{document}